\documentclass[12pt,preprint]{aastex}

\newcommand{\x}{\ensuremath{\times}}
\newcommand{\arc}{\ensuremath{^{\prime\prime}}}
\newcommand{\none}{\ensuremath{^{-1}}}
\shorttitle{Cooling Active Region Loops}
\shortauthors{Winebarger \& Warren}

\begin{document}


\title{Cooling Active Region Loops Observed With SXT and {\it TRACE}}

\author{Amy R. Winebarger \altaffilmark{1} \and Harry P. Warren}
\affil{E. O. Hulburt Center for Space Research, Code 7670, Naval
       Research Laboratory, Washington, DC 20375; winebarger@nrl.navy.mil, hwarren@nrl.navy.mil}
\altaffiltext{1}{Also at George Mason University; 4400 University Drive; Fairfax, VA 22030}


\begin{abstract}
An Impulsive Heating Multiple Strand (IHMS) Model is able to reproduce the observational
characteristics of EUV ($\sim$ 1\,MK) active region loops.  This model
implies that some of the loops must reach temperatures where X-ray 
filters are sensitive ($>$ 2.5\,MK) before they cool to EUV temperatures.  Hence, some bright EUV loops must 
be preceded by bright X-ray loops.  Previous analysis of X-ray and EUV
active region observations, however, have concluded that EUV loops are not preceded by X-ray loops.
In this paper, we examine two active regions observed in both X-ray and EUV 
filters and analyze the evolution of five loops over several hours.  These loops first appear bright in the
X-ray images and later appear bright in the EUV images.  The delay between the appearance of the loops
in the X-ray and EUV filters is as little as 1 hour and as much as 3 hours.
All five loops appear as single ``monolithic'' structures in the
X-ray images, but are resolved into many smaller structures in the (higher resolution) EUV images.
The positions of the loops appear to shift during cooling implying that the
magnetic field is changing as the loops evolve.  There is no correlation
between the brightness of the loop in the X-ray and EUV filters meaning a bright
X-ray loop does not necessarily cool to a bright EUV loop and vice versa.
The progression of the loops from X-ray images to EUV images and the observed substructure
is qualitatively consistent with the IMHS model.
\end{abstract}



\section{Introduction}

The source of coronal heating remains one of the most significant unknowns in 
solar physics.  One way to discriminate among various coronal heating theories is to reconcile 
observations of the solar corona with theoretical predictions.  For instance, hot X-ray loop 
observations tend to support that the energy release in the corona is steady
(e.g., \citealt{klimchuk1995,schrijver2004}), while EUV observations of coronal loops cannot be 
reproduced with models based on steady heating (e.g., \citealt{aschwanden2001,winebarger2003}).  Though the lifetimes of the
EUV loops are long compared to the cooling time, the intensities of the loops are too bright by several 
orders of magnitude to be able to agree with static, gravitationally stratified loops.  

\cite{warren2002b} suggested that a bundle of impulsively heated
strands could emulate the enhanced intensities and long lifetimes 
that are characteristic of EUV active region loops.  
We call this the Impulsively Heated Multiple Strand (IHMS) Model.
(We use the term ``strand'' to be the largest flux tube in which the
plasma is approximately uniform on a cross section and the term ``loop" to be
an identifiable coherent structure in an observation.  Generally, we assume that
an observed loop is a bundle of strands.)  
This heating scenario is based on the nanoflare heating mechanism suggested
by Parker, i.e., energy is released impulsively through reconnection at tangential
discontinuities in the magnetic field \citep{parker1988}.
Unlike previous implementations of the nanoflare heating model where energy
release it thought to be random in space and time and repetitive in single strands
(e.g., \citealt{cargill1997, cargill2004}), however, the IMHS model suggests that
the heating is more ordered.  A single strand is heated only once and the
strands within a bundle are heated sequentially.  The sequential heating of the
strands is empirically motivated by the extended lifetime of the loop, but is 
similar to the sequential heating during solar flares  
(e.g., \citealt{forbes1996}).

An impulsively heated strand can have densities much
larger than the densities of a loop in hydrostatic equilibrium as it cools
through typical EUV temperatures (e.g., \citealt{winebarger2004}).  A 
bundle of sequentially heated strands extends the lifetime of the observed loop.
Taken together, these characteristics can recreate the long lifetimes
and large intensities associated with the EUV coronal loops.  Because the
strands are cooling, the IHMS model predicts that the loops would be observed
in filters sensitive to hotter plasma before appearing in filters sensitive to cooler plasma.
A recent study found several examples of EUV loops that appeared in the
hotter EUV filters before the cooler EUV filters and hence were consistent with IHMS model
predictions \citep{winebarger2003a,warren2003}.    The IHMS model further
predicts broad differential emission measures (DEMs) along the loop and predominant
downflows at $\sim 1$ MK temperatures.  Broad DEMs \citep{schmelz2001} and downflows \citep{winebarger2002b} 
have been associated with the EUV loops.

To reproduce the observed EUV intensities, the strands must
sometimes be heated to large temperatures \citep{warren2002b,warren2003} implying that
some of the loops may first be observed in X-ray filter images.
However, previous analysis of coordinated active region observations made with
{\it Yohkoh}/SXT (sensitive to temperatures $> 2.5$ MK) and 
{\it TRACE} have concluded that EUV loops are not preceded by X-ray loops
(\citealt{nitta2000,schmieder2004}).  If, indeed, 
X-ray loops do not precede EUV loops, the maximum temperature of the strands would be
severely restricted and the ability of the IHMS model to
reproduce the large intensities observed in the EUV loops would be called into question.

The purpose of this paper is to determine if some EUV loops are preceded by X-ray loops.  
We examine two active regions observed with {\it Yohkoh}/SXT and
{\it TRACE} for several hours.
We follow the evolution of five loops that brighten first in SXT images, then
as much as 3 hours later, brighten in {\it TRACE} images.  These loops appear as
single structures in the SXT images, but are resolved into multiple structures in
the {\it TRACE} images.  The positions of the
loops shift during their evolution implying the magnetic field changes.
There is no correlation
between the brightness of the loop in the X-ray and EUV filters meaning a bright
X-ray loop does not necessarily cool to a bright EUV loop and vice versa.


\section{Data and Analysis}

In this section, we examine data from the {\it Transition Region and Coronal Explorer (TRACE)}
and the Soft X-ray Telescope (SXT) flown aboard the \textit{Yohkoh}
satellite to determine if loops observed by {\it TRACE} are preceded
by loops observed by SXT.  {\it TRACE}, launched 1998 April 2, has three EUV filters; 
we consider the 171\,\AA\ filter which is sensitive to the Fe~IX/X lines formed at $\sim 1$\,MK in
this study.  The images are projected on a 1024 \x 1024 CCD detector with each pixel having 
a 0.5\arc\ resolution resulting in a maximum field of view of 512\arc~\x~512\arc.
The instrument has been described in detail by \cite{handy1999},
\cite{schrijver1999}, and \cite{golub1999}.  
\textit{Yohkoh} was launched 1991 August 30 and operated until
contact was lost during a deep eclipse on 2001 December 14. The Soft
X-Ray Telescope (SXT) on {\it Yohkoh} has several focal plane filters;
we use the thin Al
filter (Al.1) and the Al, Mg, Mn, C ``sandwich'' filter (AlMg)
which are sensitive to plasmas with temperatures greater than
2.5\,MK and the Be and thick Al filters which are sensitive to hotter 
plasma in this study.  SXT has a a nominal spatial resolution of about 5\arcsec\ (2\farcs46
pixels).  The SXT CCD is 
$1024\times1024$, which allows for observations of the full solar
disk at half resolution and observations of active regions at full resolution. 
During the {\it Yohkoh} 96 minute orbit, the satellite is eclipsed by the Earth for
approximately 45 minutes causing significant gaps in the SXT data.
The SXT instrument is described in detail by \cite{tsuneta1991}.

The two active regions considered in this study are AR 8471 observed on 1999 February 27,
and AR 9017 observed on 2000 May 31.  The {\it TRACE} observations 
included the 171\,\AA\ filter.  All images with more than 100 cosmic ray hits
were removed from the study.  To improve signal-to-noise, the remaining {\it TRACE} images 
were aligned and summed at 5 minute intervals.
The SXT observations included partial frame, full
resolution images taken of each active region with the Al.1 and AlMg filters, as well 
as full disk images taken at half resolution.  The SXT observations of AR 8471 
included the Be filter, while the observations of AR 9017 included the thick Al filter.

To align the SXT and {\it TRACE} data, we use data from the Extreme
ultraviolet Imaging Telescope (EIT) flown aboard {\it SOHO} \citep{delaboudiniere1995}.  EIT provides full disk
EUV images of the Sun in filters similar to the EUV filters on {\it TRACE}.  
Limb fitting procedures were used to determine the precise pointing of the full disk 
EIT and SXT images.  The {\it TRACE} pointings were then corrected by aligning
the {\it TRACE} images to partial frames of the EIT images.  We believe the alignment
of the {\it TRACE} and SXT images to be within two SXT half-resolution pixels ($\sim 10$\arc).
EIT images were also used to  provide additional information on AR 8471 before {\it TRACE}
began observing the active region at 12:00~UT.

\begin{figure*}[t!]
\centerline{
\resizebox{18cm}{!}{\rotatebox{90}{\includegraphics[1.25in,1.25in][6in,9.75in]{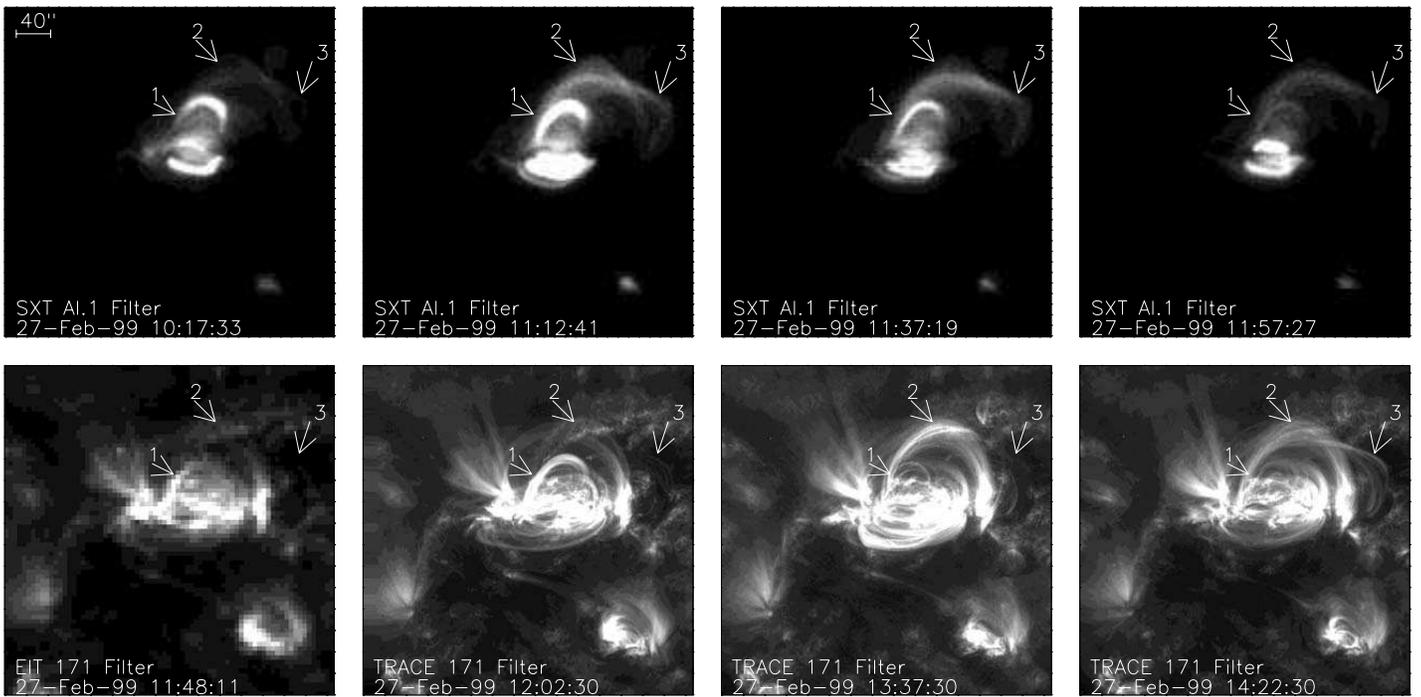}}}}
\caption{The evolution of three loops is shown with sample SXT Al.1 filter images, an EIT
171\,\AA\ filter image, and {\it TRACE} 171\,\AA\ filter images.  All images are scaled linearly and all images from 
the same filter are scaled the same.  
\label{fig:sxttrace}}
\end{figure*}

\begin{figure*}[t!]
\centerline{
\resizebox{18cm}{!}{\rotatebox{90}{\includegraphics[1.25in,1.25in][7in,9.75in]{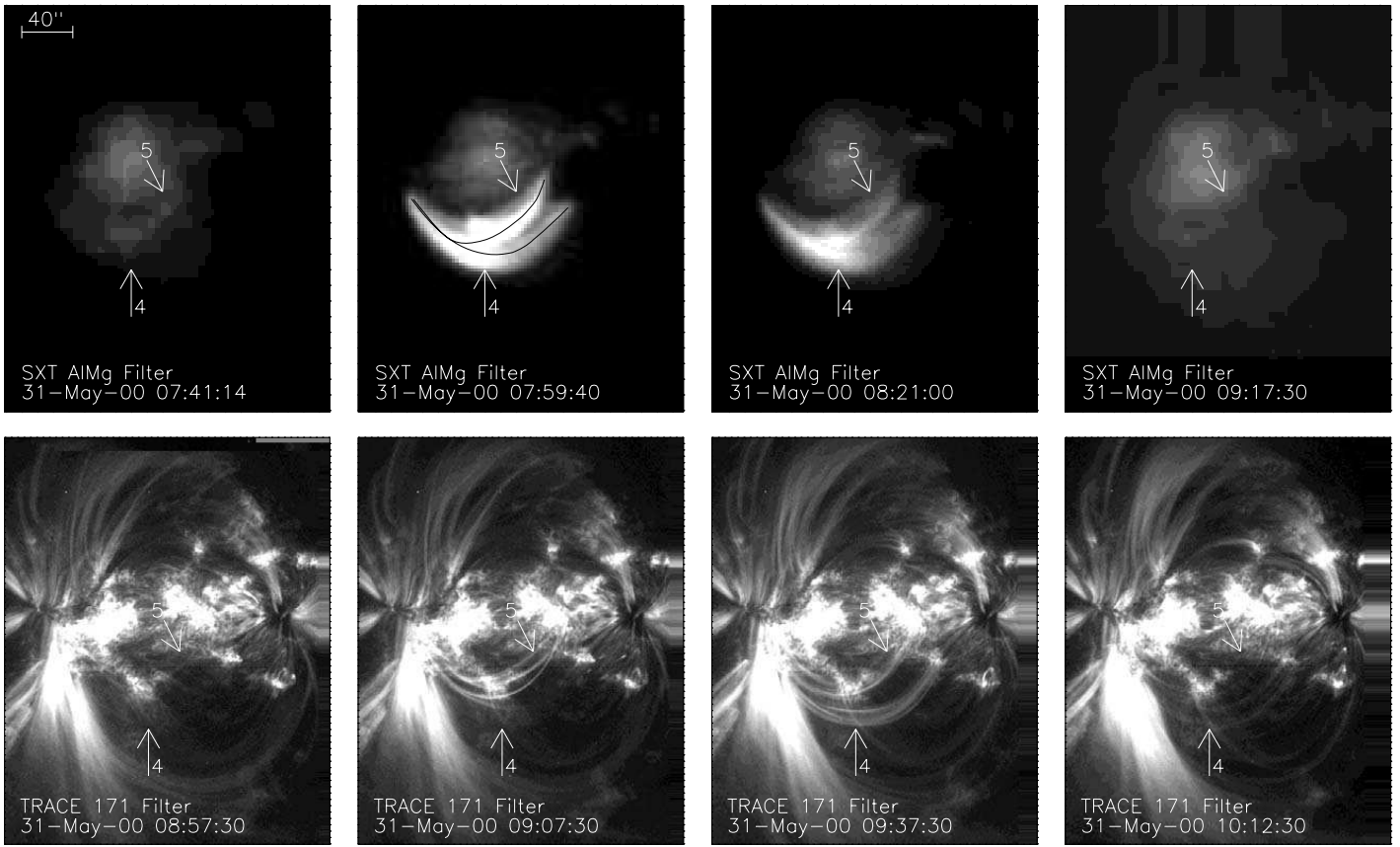}}}}
\caption{
The evolution of two loops is shown with sample SXT AlMg filter images and {\it TRACE} 171\,\AA\ images.
All images are scaled linearly and all images from the same filter are scaled the same.
\label{fig:sxttrace_000531}}
\end{figure*}

Five distinct loops were visually identified in the SXT filter images of the active regions.
On 1999 February 27, Loops 1, 2 and 3, shown in Figure~\ref{fig:sxttrace}, can be identified in the 10:17~UT SXT image (the
last image available from that orbit) but are substantially brighter at 11:12~UT.  It appears 
that Loop 1 changes orientation slightly between the 10:17~UT image and 11:12~UT image.  It is 
unclear if it is the same loop or a different loop in the same location.  The loops have dimmed in SXT by 11:57~UT
and are completely absent from images taken in the next {\it Yohkoh} orbit. 
Loop 1 may be brightening in the EIT 11:48~UT image and is definitely bright
in the first {\it TRACE} image of the active region at 12:02~UT.  The two remaining loops brighten in {\it TRACE}
substantially  later.  On 2000 May 31, Loops 4 and 5, shown in Figure~\ref{fig:sxttrace_000531}, are absent in
the 7:41~UT SXT image, but bright by 7:59~UT.  The loops fade from SXT sometime between 8:21 and 9:17~UT
during the {\it Yohkoh} eclipse.  These loops begin to appear in the {\it TRACE} filter images at 9:07~UT and
are gone by 10:12~UT.  The projected lengths of these loops and the Gaussian widths of fits to the SXT loop intensity 
($\sigma$) are given in Table~\ref{tab:loopinfo}.

In the SXT images, all of the loops appear to be monolithic structures.  For instance,
consider Loop 1 shown in Figure~\ref{fig:filament_plot}.  SXT and {\it TRACE} images of
Loop 1 are shown in the left panels.  The section of Loop 1 within the envelope
and between the cross lines is extracted and background subtracted using the method
discussed in \cite{winebarger2003a}.  The cut out from the 
SXT Al.1 image taken at 11:12~UT is shown along with a line plot of intensity averaged
along the loop.  All other images and line plots are taken from {\it TRACE} images.
The SXT intensity was fit with a Gaussian (shown with a dashed line).  The
same Gaussian function, scaled to be consistent with the {\it TRACE} intensities is 
shown in all other line plots.  The Gaussian width of the SXT loop is 1.7 SXT pixels
or 4.2\arc.  {\it TRACE} resolves several loops within what appears to be a single 
loop in the SXT data.   We fit the {\it TRACE} intensities at each time with multiple Gaussian functions
(shown with dash-dot lines); the Gaussian widths of the resulting fits in {\it TRACE} pixels are given 
in the figure.  The smallest measured width
was 4.0 {\it TRACE} pixels or 2.0\arc.   Most of the loops, however, are blended and difficult to
distinguish individually in single frames or to track from frame to frame.

A similar treatment of Loop 5 is shown in Figure~\ref{fig:filament_plot2}.   Again,
Loop 5 appears to be a single loop in the SXT image with a Gaussian width of 1.5 SXT pixels, while {\it TRACE} resolves multiple
structures.  The three {\it TRACE} images at 9:02, 9:07 and 9:12~UT, however, all image
the evolution of three distinguishable loops contained in the envelope of the SXT
loop.  The smallest Gaussian width of these loops is 2.9 {\it TRACE} pixels.  
After the data gap between 9:12 and 9:37~UT, it is unclear if those
same loops exist or if {\it TRACE} is imaging other loops.

\begin{figure}[t!h!]
\centerline{
\resizebox{8cm}{!}{\includegraphics{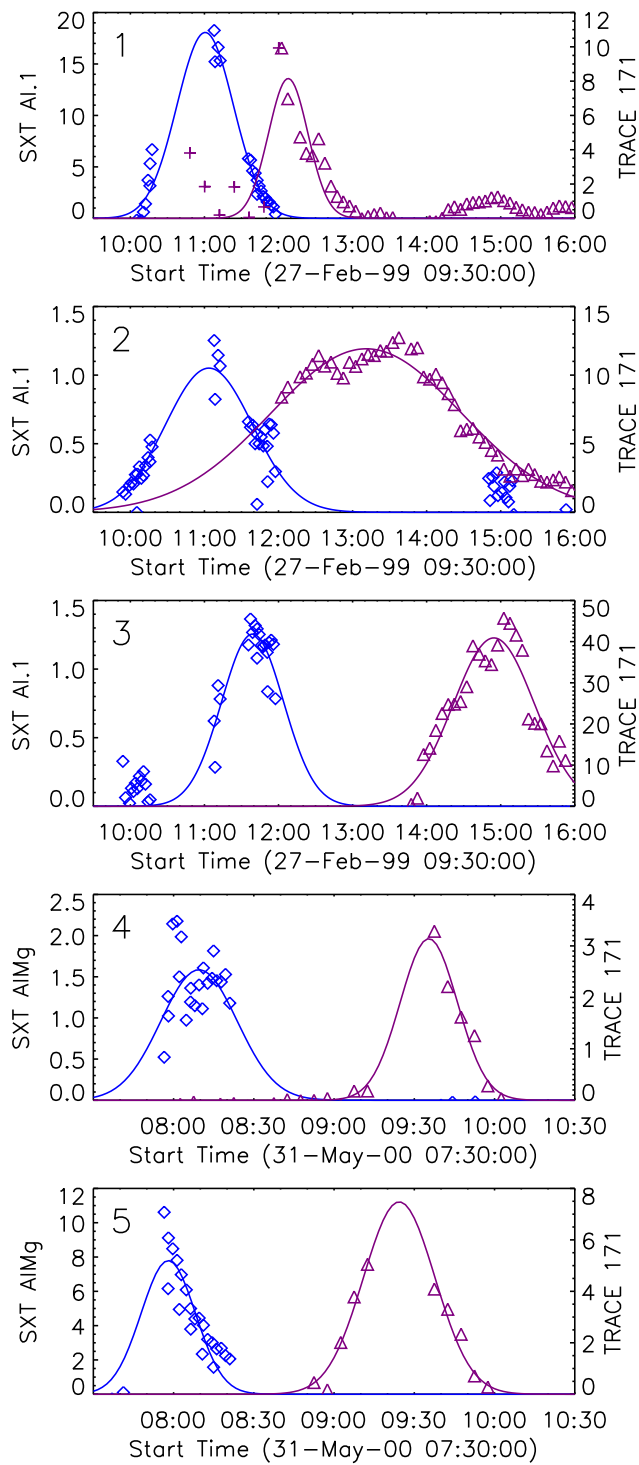}}}
\caption{The SXT Al.1 or AlMg filter intensities are shown as diamonds and are in blue.  The {\it TRACE} 171\,\AA\ intensities
are shown as triangles and are in purple.  The EIT data points are shown as crosses (only in Loop 1).  The 
Gaussian fits to the lightcurves are shown as solid lines.
The units of the SXT intensities are $10^3$\,DN~s\none and the units of the {\it TRACE} intensities are 10\,DN~s\none.
\label{fig:lc_comp}}
\end{figure}

We extracted regions of the five loops and subtracted the background.  We chose the regions of the loops
that were most secluded from other bright structures to aid in background subtraction. The extracted regions of
Loops 1 and 5 are shown in Figure~\ref{fig:filament_plot} and \ref{fig:filament_plot2}.
Regions of similar size were selected  for Loops 2, 3, and 4.  We computed lightcurves
for each filter and each loop by averaging the data along the loop within the selected region and summing the data across the
loop similar to the method discussed in \cite{winebarger2003a} resulting in the DN~s\none\ for each
loop as a function of time.
The SXT Al.1 filter and {\it TRACE} 171\,\AA\ filter lightcurves for the three loops in AR 8471 and  the SXT  AlMg
filter and {\it TRACE} 171\,\AA\ filter lightcurves for the two loops in AR 9017
are shown in Figure~\ref{fig:lc_comp}.
For Loop 1, additional points were added to the lightcurve before 12:00~UT by extracting the
same region from aligned EIT data and scaling the measured intensities to match the {\it TRACE} 171\,\AA\ intensities
at the 12:00 time.  The lightcurves were fit with Gaussian functions, shown as solid lines.
Note that because of the data gaps and the absence of {\it TRACE} data before 12:00~UT on 1999 Feb 27, 
the fits to the lightcurves are not well constrained and the information derived from them 
(the delay and lifetime)  are approximations.  
The delay between the peak of the lightcurves in the two filters is approximately 
67 minutes for Loop 1, 130 minutes for Loop 2, 200 minutes for Loop 3, 86 minutes for Loop 4
and 86 minutes for Loop 5.  The lifetimes of
the loops in each filter are assumed to be the full width half maximum of the fits and 
are given in Table~\ref{tab:loopinfo}.

Figure~\ref{fig:lc_comp} and Table~\ref{tab:loopinfo} give information on the lightcurves in the
most sampled SXT filters.  In Table~\ref{tab:lcinfo}, we give the maximum intensities of the loops
in all available filters.  These values represent the maximum
observed values and are not based on a fit.  We give these numbers in two separate ways.  The first
is the intensity summed across the loop, represented as $I_{filter}^{total}$.  These values are
what is plotted in Figure~\ref{fig:lc_comp} and are compatible with our previous analysis (\citealt{winebarger2003a}).
We also provide this total value divided by $\sqrt{2\pi}\sigma$ where $\sigma$ is the width of the loop in 
either {\it TRACE} or SXT pixels found from the SXT filter images and given in Table~\ref{tab:loopinfo}. 
This value provides an estimate of the typical observed countrates per instrument pixel of the loop.  For instance,
the typical countrates for Loop 1 go from 4,200 DN s$^{-1}$ pixel$^{-1}$ in the SXT Al.1 filter images to 
4.7 DN s$^{-1}$ pixel$^{-1}$ in {\it TRACE} 171\,\AA\ filter images. Loop 1 is also fairly bright in the SXT Be filter images.

While the range of typical {\it TRACE} intensities is relatively narrow (1.7-4.7 DN s\none pixel\none), the range
of SXT filter intensities is quite broad (230-5,100 DN s\none pixel\none\ in the SXT Al.1 filter).
Furthermore, there is little correlation between the {\it TRACE} 171\,\AA\ intensities and the
SXT intensities.  For instance, Loops 1 and 2  and Loops 4 and 5 both have approximately the same intensity in the 
{\it TRACE} 171\,\AA\ filter, but have an order of magnitude difference in intensity in both the SXT filters.  
Loops 2 and 3 have approximately the same intensities in the SXT Al.1 filter, but there is a factor of
2 difference in their {\it TRACE} intensities.  There is also little correlation between the delay of the
appearance of the loop in the different filters (given in Table~\ref{tab:loopinfo}) and the intensities.
Loops 4 and 5 have the same delay, but significantly different intensities in the SXT filters.
There is a correlation between the intensity of the loop in the hotter SXT filters (Be or
thick Al) and the ratio of the intensities in the cooler SXT filter to the {\it TRACE} 171 \AA\ filter.


\section{Discussion}

In this paper, we have shown the evolution of five cooling loops.
These loops first appear bright in SXT filter images (sensitive to plasma $> 2.5$\,MK) and
then appear bright in {\it TRACE} filter images (sensitive to plasma $\sim 1$\,MK).
All loops appear as single, ``monolithic'' structures in the SXT images with 
Gaussian widths on the order of the resolution of the SXT instrument.  These loops are 
resolved into several loops, each evolving independently, by {\it TRACE}.  The widths of
the {\it TRACE} loops are at least 3 {\it TRACE} pixels.  These observations are qualitatively consistent with the
predictions of the Impulsively Heated Multiple Strand Model.

The fact that there is no correlation between the brightness of the loops in SXT and {\it TRACE}
filters demonstrates the non-linearity of the physics driving the plasma properties.  
It is impossible to say that a bright loop in {\it TRACE} must have been preceded by a bright loop
in SXT or vice versa.  Instead the evolution of the loops depends on the loop length, energy input, elemental
abundances of the plasma, etc.  
Ideally, to understand better the evolution of these
loops, we would use hydrodynamic models of each loop's length and attempt to 
match the observed lightcurves exactly.   
Problems with the current data set such as the long data gaps caused by the {\it Yohkoh} orbit
make quantitative modeling difficult.  Furthermore, we were not able
to resolve the loop length and geometry.  The loop lengths were not well
constrained by the geometric arguments used in previous analysis (see \citealt{winebarger2003a}).
It is difficult to make a conclusive statement concerning the
correlation between the intensity of the loop in the hotter SXT filters 
and the ratio of the intensity in the cooler SXT filter to the {\it TRACE} 171 \AA\ filter.
It may be an important metric to aid in quantitative models.

\begin{figure}[t!]
\centerline{
\resizebox{8cm}{!}{\includegraphics{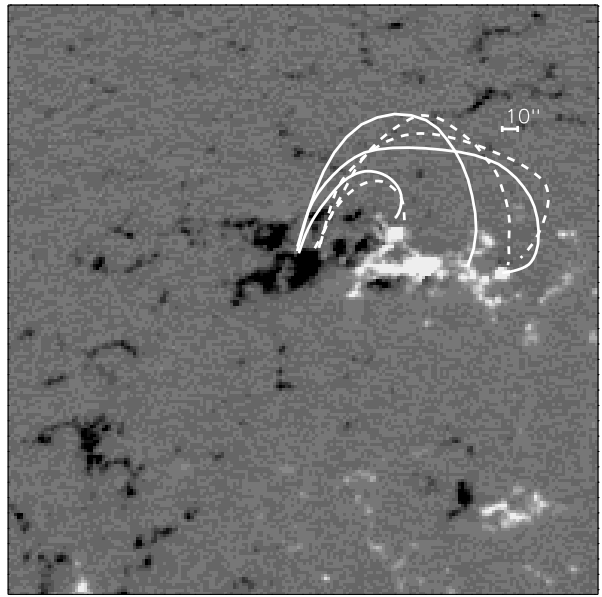}}}
\caption{The relative positions of Loops 1, 2, and 3 determined from {\it TRACE} (solid) and SXT (dashed) images
shown on a magnetogram from MDI.
\label{fig:nitta}}
\end{figure}

\cite{nitta2000} and \cite{schmieder2004} have examined SXT and {\it TRACE} active
region observations for evidence of cooling loops.  Though they observed that
some loops would brighten in {\it TRACE} images in similar locations to where loops
had previously brightened in SXT images, they concluded they were different loops
due to the positional shifts.  The five loops presented in this paper appear to shift as well.
Figure~\ref{fig:nitta} shows the relative positions of Loops 1, 2 and 3
determined from the {\it TRACE} data (solid lines) and SXT data (dashed lines) on a
magnetogram from the Michelson Doppler Imager (MDI) flown aboard {\it SOHO}.  
The positions of the loops were found by ``clicking on'' the loop in the different
images.   The positions of the loops are most uncertain near the footpoints.  In the
{\it TRACE} images, many bright structures obscure the loops near the footpoints
while the SXT filters are only sensitive to coronal plasma and hence do not 
image the true footpoint of the loop.
The alignment of the images, which we believe is good to 10\arc, cannot account for
all the differences in the coronal part of the loops' positions. 
Magnetic field changes could account for the shift in position, similar to the
magnetic motions observed during solar flares (e.g., \citealt{forbes1996}).

\begin{figure*}[t!]
\centerline{
\resizebox{18cm}{!}{\rotatebox{90}{\includegraphics[0in,1.5in][6.5in,10in]{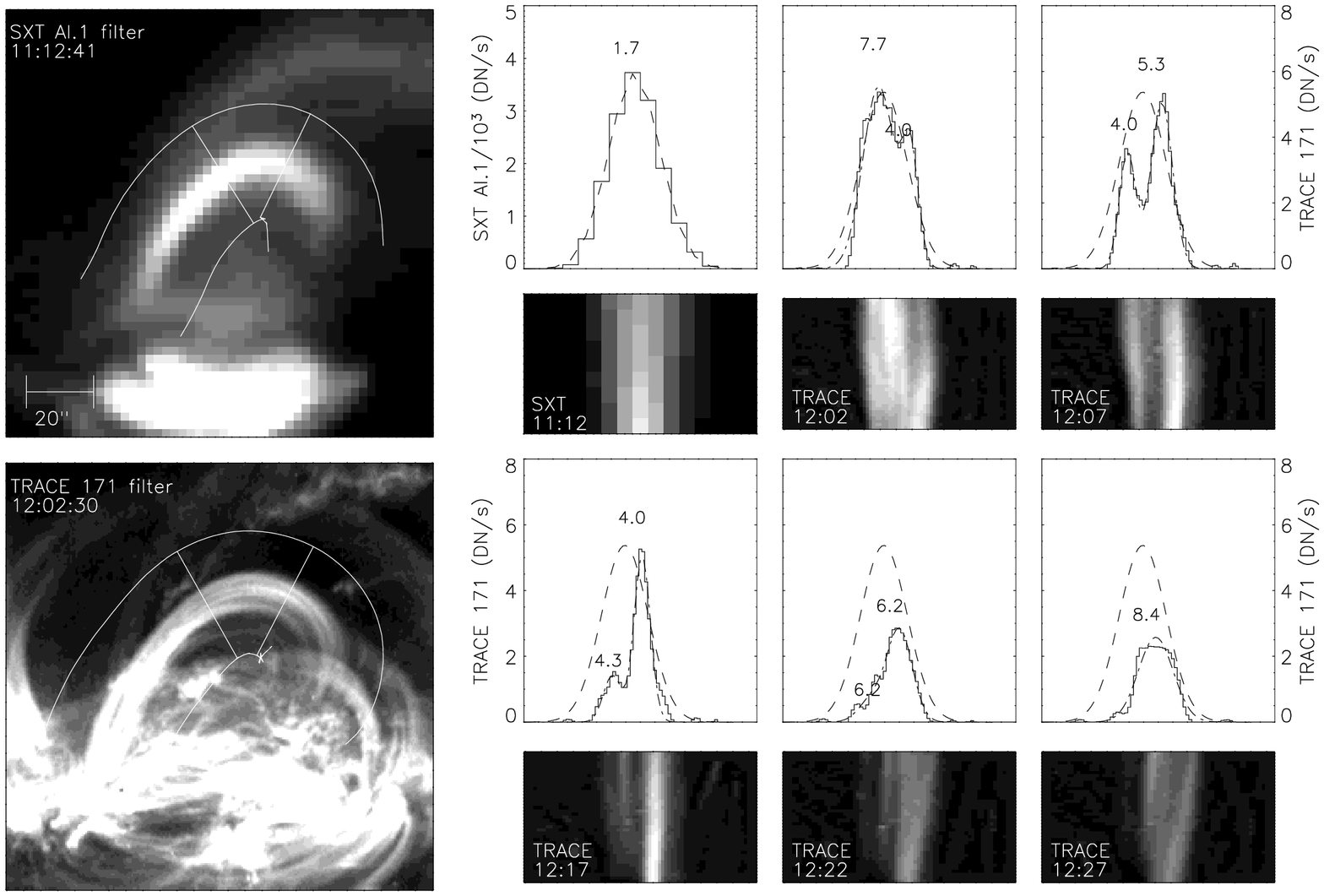}}}}
\caption{SXT Al.1 filter and {\it TRACE} 171\,\AA\ filter images of Loop 1 are shown on the left.  The
evolution of the area of the loop within the envelope and between the cross lines is
shown in the smaller images.  The line plots show the intensity summed along the loop.
\label{fig:filament_plot}}
\end{figure*}

\begin{figure*}[t!]
\centerline{
\resizebox{18cm}{!}{\rotatebox{90}{\includegraphics[0in,1.5in][6.5in,10in]{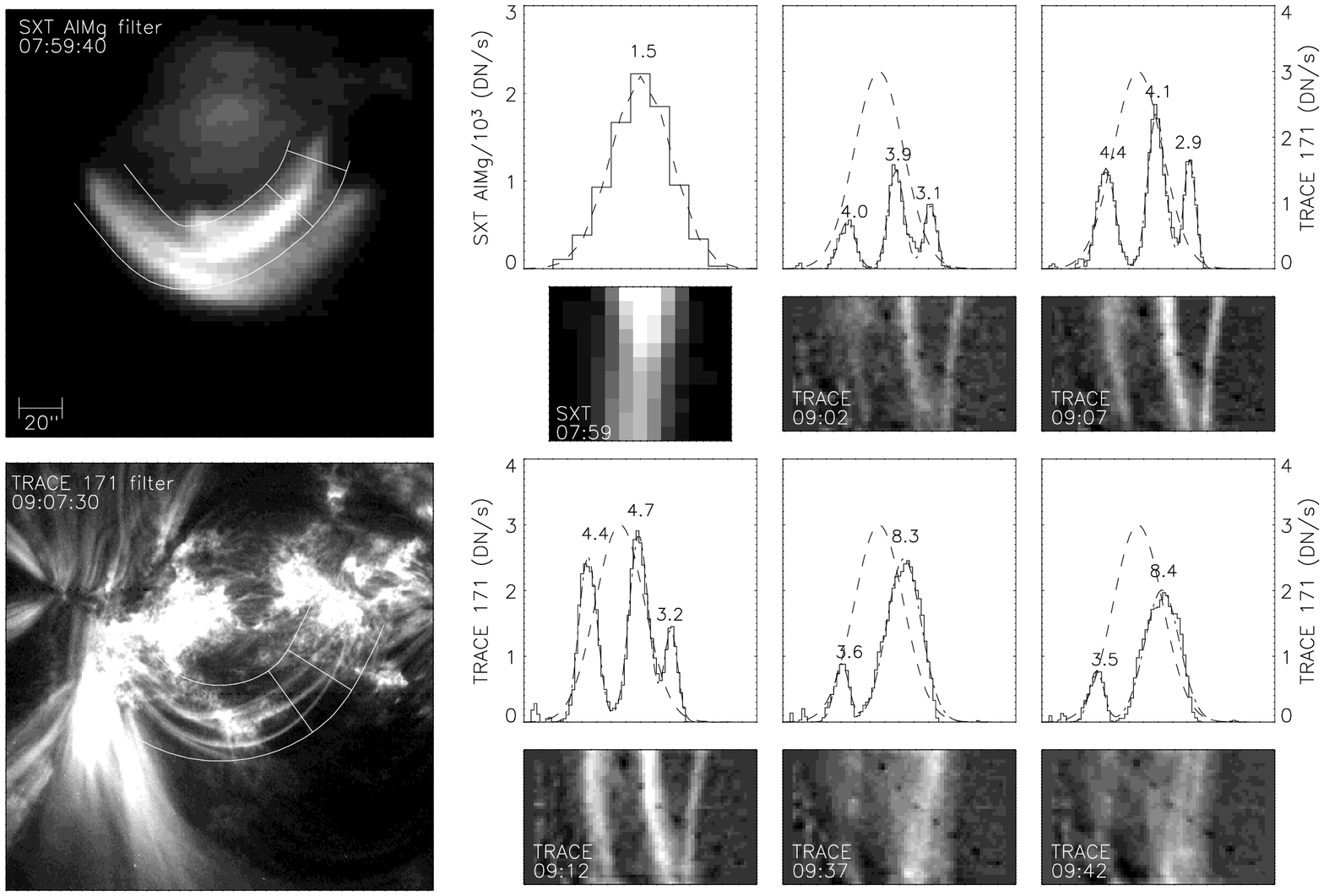}}}}
\caption{SXT AlMg filter and {\it TRACE} 171\,\AA\ filter images of Loop 5 are shown on the left.  The
evolution of the area of the loop within the envelope and between the cross lines is
shown in the smaller images.  The line plots show the intensity summed along the loop.
\label{fig:filament_plot2}}
\end{figure*}

The IHMS model is similar to the scenario of energy release during  solar flares.
During large solar flares, however, the loops brighten in a systematic way,
as the magnetic field sweeps through the reconnection region.
Images of Loops 1 and 5 shown in Figures~\ref{fig:filament_plot} and \ref{fig:filament_plot2}, 
however, show that there is no systematic way in which the loops within the bundle brighten. 
Though the IHMS model is motivated by the magnetic reconnection models, it does not 
exclude other impulsive heating models where strands within a loop are heated
sequentially.  For instance, \cite{ofman1998} suggests that the heating due to resonant absorption of
Alfv\'{e}n waves has similar spatial and temporal characteristics as predicted by
the IHMS model.

We demonstrate that what appears to be single loop in SXT images is actually a bundle of loops.
Some of the individual loops in {\it TRACE} appear to be resolved,
such as the three loops comprising ``Loop 5" at 9:02, 9:07, and 9:12~UT shown in
Figure \ref{fig:filament_plot2} because their measured width exceeds the resolution of
the {\it TRACE} instruments (2 pixels).  Most of the loops, though, are blended together
and difficult to distinguish.  In a future investigation, we will determine if the
individual loops within the bundle observed with {\it TRACE} are consistent with single 
temperatures and densities and hence are monolithic strands or if they, too, must
contain a bundle of strands to explain their properties.

This paper has addressed moderate length active region loops that are completely visible (footpoint to
apex) in EUV images (see \citealt{aschwanden2000a} for a description of a large sample of these loops).  
It does not address other types of active region loops, such as the extended ``fan''
structures at the edges of active regions which are only visible in the EUV near their 
footpoints or the short, hot loops in active region cores that are bright in X-ray images
and whose footpoints form the bright reticulated pattern called ``moss'' in EUV data. 
\cite{antiochos2003} argue that the the moss loops must be heated steadily because the
loops remain bright in X-ray images and never appear to cool through EUV images.
\cite{cargill2004} suggests that the heating is dynamic, but that multiple heating events within a strand 
keep the plasma at hotter temperatures.  If future studies of these
other structures confirm that the temporal characteristics of the energy deposition
is markedly different than suggested by the IHMS model, it could imply that there are
multiple mechanisms that release energy into the corona.


\acknowledgments

This research was funded by a grant from NASA's Solar and Heliospheric Physics program.



\begin{deluxetable}{lccccccc}
\footnotesize
\tablecaption{Loop information.}
\tablewidth{0pt}
\tablehead{Loop & Date & AR Number & Length & $\sigma$ & Delay &  FWHM - SXT &  FWHM - {\it TRACE} \\
& & & (Mm) &  (arcsec) & (minutes) & (minutes)&  (minutes) }
\startdata
1 & 1999 Feb 27 & 8471 &  79 &  4.2 &  67 & 54 & 38 \\
2 & 1999 Feb 27 & 8471 & 178 &  5.4 & 127 & 83 & 182 \\
3 & 1999 Feb 27 & 8471 & 190 &  5.1 & 196 & 59 & 78 \\
4 & 2000 May 31 & 9017 & 109 &  3.9 & 86  & 34 & 25 \\
5 & 2000 May 31 & 9017 &  98 &  3.9 & 86  & 24 & 31 \\
\enddata
\label{tab:loopinfo}
\end{deluxetable}

\begin{deluxetable}{lcccccccccccc}
\footnotesize
\tablecaption{Lightcurve information.}
\tablewidth{0pt}
\tablehead{Loop & $I_{Be}^{total}$ & $I_{Be}^{pix}$ & $I_{Al12}^{total}$ &  $I_{Al12}^{pix}$ &
	$I_{AlMg}^{total}$ & $I_{AlMg}^{pix}$ &  $I_{Al.1}^{total}$ &
	$I_{Al.1}^{pix}$ & $I_{171}^{total}$ & $I_{171}^{pix}$}
\startdata
1  & $150$ & $35$  &     &     & $9,300$  & $2,200$  & $18,000$  & $4,200$  &  $99$  & 4.7 \\
2  & $10$  & $1.9$ &     &     & $720 $   & $130$    & $1,200$   & $230$    & $130$  & 4.7 \\
3  & $40$  & 7.7   &     &     & $620$    & $120$    & $1,400$   & $260$    & $46$   & 1.8 \\
4  &       &       & 170 & 43  & $2,200$  & $540$    & $3,600$   & $890$    & $33$   & 1.7 \\
5  &       &       & 730 & 180 & $11,000$ & $2,600$  & $21,000$  & $5,100$  & $50$   & 2.6 \\
\enddata
\label{tab:lcinfo}
\end{deluxetable}


\clearpage

\clearpage

\clearpage

\clearpage

\clearpage

\clearpage


\begin{thebibliography}{}

\bibitem[\protect\citeauthoryear{{Antiochos} et~al.}{{Antiochos}
  et~al.}{2003}]{antiochos2003}
{Antiochos}, S.~K., {Karpen}, J.~T., {DeLuca}, E.~E., {Golub}, L.,  \&
  {Hamilton}, P. 2003, \apj, 590, 547

\bibitem[\protect\citeauthoryear{{Aschwanden}, {Nightingale}, \&
  {Alexander}}{{Aschwanden} et~al.}{2000}]{aschwanden2000a}
{Aschwanden}, M.~J., {Nightingale}, R.~W.,  \& {Alexander}, D. 2000, \apj, 541,
  1059

\bibitem[\protect\citeauthoryear{{Aschwanden}, {Schrijver}, \&
  {Alexander}}{{Aschwanden} et~al.}{2001}]{aschwanden2001}
{Aschwanden}, M.~J., {Schrijver}, C.~J.,  \& {Alexander}, D. 2001, \apj, 550,
  1036

\bibitem[\protect\citeauthoryear{{Cargill} \& {Klimchuk}}{{Cargill} \&
  {Klimchuk}}{1997}]{cargill1997}
{Cargill}, P.~J.,  \& {Klimchuk}, J.~A. 1997, \apj, 478, 799

\bibitem[\protect\citeauthoryear{{Cargill} \& {Klimchuk}}{{Cargill} \&
  {Klimchuk}}{2004}]{cargill2004}
{Cargill}, P.~J.,  \& {Klimchuk}, J.~A. 2004, \apj, 605, 911

\bibitem[\protect\citeauthoryear{{Delaboudiniere} et~al.}{{Delaboudiniere}
  et~al.}{1995}]{delaboudiniere1995}
{Delaboudiniere}, J.-P., et~al. 1995, \solphys, 162, 291

\bibitem[\protect\citeauthoryear{{Forbes} \& {Acton}}{{Forbes} \&
  {Acton}}{1996}]{forbes1996}
{Forbes}, T.~G.,  \& {Acton}, L.~W. 1996, \apj, 459, 330

\bibitem[\protect\citeauthoryear{Golub et~al.}{Golub et~al.}{1999}]{golub1999}
Golub, L., et~al. 1999, Phys. Plasmas, 6, 2205

\bibitem[\protect\citeauthoryear{{Handy} et~al.}{{Handy}
  et~al.}{1999}]{handy1999}
{Handy}, B.~N., et~al. 1999, \solphys, 187, 229

\bibitem[\protect\citeauthoryear{{Klimchuk} \& {Porter}}{{Klimchuk} \&
  {Porter}}{1995}]{klimchuk1995}
{Klimchuk}, J.~A.,  \& {Porter}, L.~J. 1995, Nature, 377, 131

\bibitem[\protect\citeauthoryear{{Nitta}}{{Nitta}}{2000}]{nitta2000}
{Nitta}, N. 2000, \solphys, 195, 123

\bibitem[\protect\citeauthoryear{{Ofman}, {Klimchuk}, \&
  {Davila}}{{Ofman} et~al.}{1998}]{ofman1998}
{Ofman}, L. and {Klimchuk}, J.~A. and {Davila}, J.~M. 1998, \apj, 493, 474

\bibitem[\protect\citeauthoryear{{Parker}}{{Parker}}{1988}]{parker1988}
{Parker}, E.~N. 1988, \apj, 330, 474

\bibitem[\protect\citeauthoryear{{Schmelz} et~al.}{{Schmelz}
  et~al.}{2001}]{schmelz2001}
{Schmelz}, J.~T., {Scopes}, R.~T., {Cirtain}, J.~W., {Winter}, H.~D.,  \&
  {Allen}, J.~D. 2001, \apj, 556, 896

\bibitem[\protect\citeauthoryear{{Schmieder} et~al.}{{Schmieder}
  et~al.}{2004}]{schmieder2004}
{Schmieder}, B., {Rust}, D.~M., {Georgoulis}, M.~K., {D{\' e}moulin}, P.,  \&
  {Bernasconi}, P.~N. 2004, \apj, 601, 530

\bibitem[\protect\citeauthoryear{{Schrijver} et~al.}{{Schrijver}
  et~al.}{2004}]{schrijver2004}
{Schrijver}, C.~J., {Sandman}, A.~W., {Aschwanden}, M.~J.,  \& {DeRosa}, M.~L.
  2004, \apj, 615, 512

\bibitem[\protect\citeauthoryear{{Schrijver} et~al.}{{Schrijver}
  et~al.}{1999}]{schrijver1999}
{Schrijver}, C.~J., et~al. 1999, \solphys, 187, 261

\bibitem[\protect\citeauthoryear{{Tsuneta} et~al.}{{Tsuneta}
  et~al.}{1991}]{tsuneta1991}
{Tsuneta}, S., et~al. 1991, \solphys, 136, 37

\bibitem[\protect\citeauthoryear{{Warren}, {Winebarger}, \&
  {Hamilton}}{{Warren} et~al.}{2002}]{warren2002b}
{Warren}, H.~P., {Winebarger}, A.~R.,  \& {Hamilton}, P.~S. 2002, \apjl, 579,
  L41

\bibitem[\protect\citeauthoryear{{Warren}, {Winebarger}, \& {Mariska}}{{Warren}
  et~al.}{2003}]{warren2003}
{Warren}, H.~P., {Winebarger}, A.~R.,  \& {Mariska}, J.~T. 2003, \apj, 593,
  1174

\bibitem[\protect\citeauthoryear{{Winebarger} \& {Warren}}{{Winebarger} \&
  {Warren}}{2004}]{winebarger2004}
{Winebarger}, A.~R.,  \& {Warren}, H.~P. 2004, \apjl, 610, L129

\bibitem[\protect\citeauthoryear{{Winebarger}, {Warren}, \&
  {Mariska}}{{Winebarger} et~al.}{2003a}]{winebarger2003}
{Winebarger}, A.~R., {Warren}, H.~P.,  \& {Mariska}, J.~T. 2003a, \apj, 587, 439

\bibitem[\protect\citeauthoryear{{Winebarger}, {Warren}, \&
  {Seaton}}{{Winebarger} et~al.}{2003b}]{winebarger2003a}
{Winebarger}, A.~R., {Warren}, H.~P.,  \& {Seaton}, D.~B. 2003b, \apj, 593, 1164

\bibitem[\protect\citeauthoryear{{Winebarger} et~al.}{{Winebarger}
  et~al.}{2002}]{winebarger2002b}
{Winebarger}, A.~R., {Warren}, H.~P., {van Ballegooijen}, A., {DeLuca}, E.~E.,
  \& {Golub}, L. 2002, \apj, 567, L89

\end{thebibliography}
\end{document}